\documentclass[dvipdfmx]{pasj01}

\begin{document} 
\Received{2016/03/29}
\Accepted{2016/05/12}

\title{Gas-to-dust ratio in massive star-forming galaxies at $z\sim1.4$}

\author{Akifumi \textsc{seko}\altaffilmark{1}}%
\altaffiltext{1}{Department of Astronomy, Kyoto University, Kitashirakawa-Oiwake-Cho, Sakyo-ku, Kyoto 606-8502}
\email{seko@kusastro.kyoto-u.ac.jp}

\author{Kouji \textsc{ohta}\altaffilmark{1}}

\author{Kiyoto \textsc{yabe}\altaffilmark{2,3}}
\altaffiltext{2}{National Astronomical Observatory of Japan, 2-21-1 Osawa, Mitaka, Tokyo 181-8588}
\altaffiltext{3}{Kavli Institute for the Physics and Mathematics of the Universe (Kavli IPMU, WPI), 
The University of Tokyo Institutes for Advanced Study, The University of Tokyo, Kashiwa, Chiba 277-8583, Japan}

\author{Bunyo \textsc{hatsukade}\altaffilmark{2}}

\author{Yuya \textsc{aono}\altaffilmark{1}}

\author{Daisuke \textsc{iono}\altaffilmark{2,4}}
\altaffiltext{4}{The Graduate University for Advanced Studies (SOKENDAI), 2-21-1 Osawa, Mitaka, Tokyo 181-0015}


\maketitle

\begin{abstract}
We present results of $^{12}$CO($J=2-1$) observations toward four massive star-forming galaxies at $z\sim1.4$ 
with the Nobeyama 45~m radio telescope. 
The galaxies are detected with {\it Spitzer}/MIPS in 24~$\mu\mathrm{m}$, {\it Herschel}/SPIRE 
in 250~$\mu\mathrm{m}$, and 350~$\mu\mathrm{m}$ and they mostly reside in the main sequence. 
Their gas-phase metallicities derived with N2 method by using the H$\alpha$ 
and [NII]$\lambda$~6584 emission lines are near the solar value. 
CO lines are detected toward three galaxies. 
The molecular gas masses obtained are $(9.6-35)\times10^{10}~M_\odot$ 
by adopting the Galactic CO-to-H$_2$ conversion factor 
and the CO(2-1)/CO(1-0) flux ratio of 3. 
The dust masses derived with the modified blackbody model 
(assuming the dust temperature of 35~K and the emissivity index of 1.5) 
are $(2.4-5.4)\times10^{8}~M_\odot$. 
The resulting gas-to-dust ratios (not accounting for HI mass) at $z\sim1.4$ are 220-1450, 
which are several times larger than those in local star-forming galaxies. 
A dependence of the gas-to-dust ratio on the far-infrared luminosity density 
is not clearly seen. 
\end{abstract}

\section{Introduction}
The redshift around 2 is a crucial epoch to understand galaxy evolution. 
The star-formation rate (SFR) density peaked around this redshift 
(e.g., \cite{Hopk06, Mada14}), 
i.e., galaxies are considered to evolve dramatically at this epoch. 
SFRs of most of star-forming galaxies correlate well with their stellar mass 
which makes a sequence in a diagram of stellar mass versus SFR. 
The sequence is called main sequence, and is seen at each redshift 
up to at least $z\sim2.5$
(e.g., \cite{Noes07, Dadd07, Rodi10}; \authorcite{Whit12}~\yearcite{Whit12}, \yearcite{Whit14}; \cite{Spea14}). 
Since such galaxies are primary population among star-forming galaxies, 
it is important to unveil the nature of the interstellar medium (ISM) in main-sequence galaxies 
for the understanding of galaxy evolution.

\begin{figure*}
\begin{center}
\includegraphics[width=17cm]{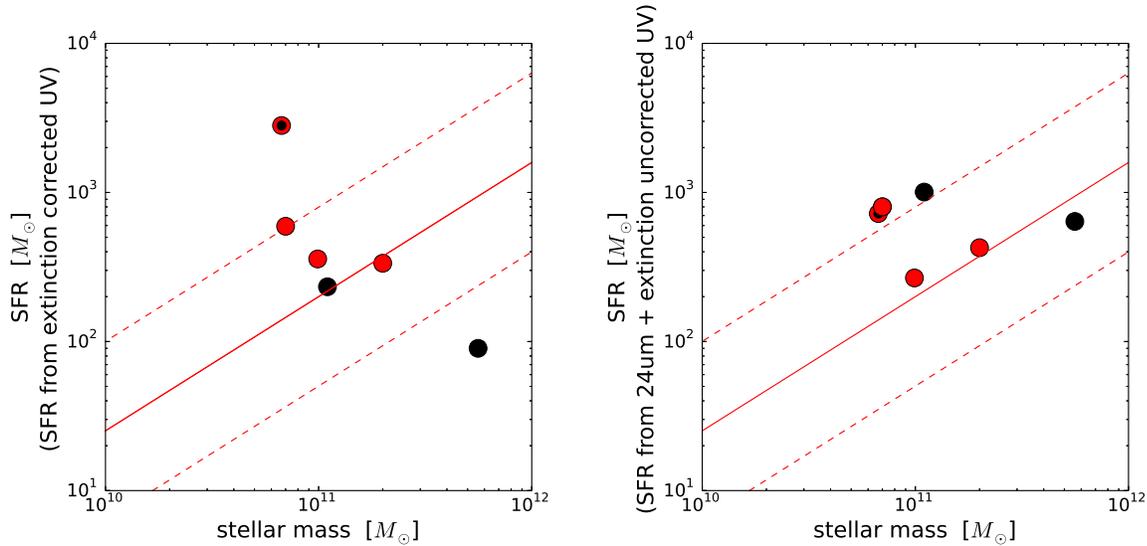} 
\end{center}
\caption{Galaxy sample in the stellar mass$-$SFR diagram. SFRs are derived from 
	extinction corrected UV luminosity densities (left), and sum of SFRs from 24~$\mu\mathrm{m}$ and 
	SFRs from extinction uncorrected UV luminosity densities (right). 
	Red and black filled circles show the galaxies in this study and \citet{Seko14}, respectively. 
	Red circle with black circle refer to the COSMOS\_9 which is one of the target galaxies in this study and \citet{Seko14}. 
	Solid line represents main sequence by \citet{Dadd10} derived for galaxies at $z\sim2$. 
	Dashed lines show the scatter of $\pm0.6$~dex.
	} \label{fig: sample}
\end{figure*}

The advent of high-sensitivity millimeter, sub-millimeter, and far-infrared telescopes enables us 
to investigate molecular gas and dust in the main sequence star-forming galaxies around $z\sim2$ 
(e.g., \cite{Dadd10, Genz10, Tacc10, Tacc13, Seko16, Elba11, Magd12b, Magn12, Magn14}). 
Recent studies of molecular gas in main sequence galaxies at $z=1-2.5$ showed 
the molecular gas mass and its fraction against stellar mass 
[$f_\mathrm{mol} = M_\mathrm{mol}/(M_\mathrm{mol} + M_\ast)$]
were significantly larger than those in local spiral galaxies 
(e.g., \cite{Dadd10, Tacc10, Tacc13, Seko16}). 
The dust mass in main sequence galaxies at $z\sim1-2$ was also found to be larger 
than that in the present-day spiral galaxies (e.g., \cite{Magd12b, Magn12, Seko16}). 

The gas-to-dust ratio in galaxies is also an important parameter that is closely related to galaxy evolution. 
In the local universe, gas-to-dust ratio is $\sim150$ at solar metallicity and depends on gas-phase metallicity; 
the ratio decreases with increasing metallicity (e.g., \cite{Lero11, Remy14}). 
In the high redshift universe, the number of main-sequence galaxies observed with both CO and dust 
with known metallicities is still very limited. 
\citet{Sain13} derived gas-to-dust ratios in lensed main-sequence galaxies at $z=1.4-3.1$ 
from the CO and dust emissions detected with IRAM Plateau de Bure Interferometer 
and {\it Herschel}/PACS and SPIRE, respectively. 
The metallicities of their galaxy sample were mostly derived 
with the R23 (by using the H$\beta$, [OII]$\lambda$~3727, and [OIII]$\lambda\lambda$~4959,5007 emission lines)
or N2 (by using H$\alpha$, [NII]$\lambda$~6584) method. 
They showed the gas-to-dust ratio was about two times larger than that in local galaxies at a fixed metallicity. 
\citet{Seko14} presented the results of CO observations of three massive star-forming galaxies 
at $z\sim1.4$ detected with MIPS and SPIRE. 
Their metallicities derived using the N2 method were near the solar value.  
Although the CO emissions from each galaxy were not detected, 
the stacked CO profile shows the emission, and the molecular gas mass was derived. 
With the average dust mass, they showed the average gas-to-dust ratio was comparable to that in local galaxies with solar metallicity. 
By using ALMA, \citet{Seko16} observed CO and dust emission of twenty main sequence galaxies with known metallicity at $z\sim1.4$. 
Almost all of them are not detected in MIPS nor SPIRE. 
The result from the stacking analysis showed that the gas-to-dust ratio was $3-4$ times 
larger than that in local galaxies at a fixed metallicity. 
The gas-to-dust ratio may depend on the infrared flux (or luminosity).
Therefore, we need to examine the gas-to-dust ratios in more galaxies detected in mid/far-infrared.

In this paper, we present $^{12}$CO($J=2-1$) observations of four massive star-forming galaxies 
with near solar metallicity at $z\sim1.4$ which are detected with {\it Spiter}/MIPS in $24~\mu\mathrm{m}$, 
{\it Herschel}/SPIRE in $250~\mu\mathrm{m}$ and $350~\mu\mathrm{m}$ to study whether 
star-forming galaxies detected in mid/far-infrared wavelength show the similar gas-to-dust ratio in local galaxies. 
Sample selection and observations and data reductions are described in Section~\ref{sec: sample}, and \ref{sec: obs and reduc}, respectively. 
We present the results, a discussion, and summary in section~\ref{sec: result}, 
section~\ref{sec: dis}, and section~\ref{sec: sum}, respectively. 
Throughout this paper, the $\Lambda$-CDM cosmology with $H_0 = 70~\mathrm{km~s^{-1}~Mpc^{-1}}$, 
$\Omega_\mathrm{M} = 0.30$, and $\Omega_\mathrm{\Lambda} = 0.70$ is adopted. 

\begin{longtable}{lccccccc}
\caption{Galaxy sample.} \label{tab: sample}
\hline              
Source & RA & Dec & $z_\mathrm{spec}$ & metallicity & $M_\ast$\footnotemark[$\ast$] & SFR\footnotemark[$\ast$, $\dagger$] & SFR\footnotemark[$\ast$, $\ddagger$] \\
 & (J2000.0) & (J2000.0) &  & [$12+\log(\mathrm{O/H)}$] & ($M_\odot$) & ($M_\odot~\mathrm{yr^{-1}}$) & ($M_\odot~\mathrm{yr^{-1}}$) \\
\endhead
  \hline
SXDS1\_13015 & \timeform{02h17m13.63s} & \timeform{-5D09'39.8''}  & 1.451 & $8.85_{-0.04}^{+0.04}$ & $2.0\times10^{11}$ 
& 335 & 426 \\
COSMOS\_9\footnotemark[$\S$]     & \timeform{10h00m08.76s} & \timeform{+2D19'02.3''} & 1.461 & $<8.68$ & $6.7\times10^{10}$ 
& 2812 & 722 \\
COSMOS\_50   & \timeform{10h01m40.28s} & \timeform{+2D33'30.9''} & 1.21 & $8.84_{-0.11}^{+0.11}$ & $7.0\times10^{10}$ 
& 593 & 801 \\
COSMOS\_53   & \timeform{10h01m36.15s} & \timeform{+2D20'04.3''} & 1.21 & $8.68_{-0.11}^{+0.11}$ & $9.9\times10^{10}$ 
& 358 & 267 \\
\hline
SXDS1\_12778\footnotemark[$\S$] & \timeform{02h19m09.45s} & \timeform{-5D09'49.0''} & 1.396 & $8.66_{-0.13}^{+0.09}$ & $5.6\times10^{11}$ 
& 90 & 639 \\
SXDS3\_80799\footnotemark[$\S$] & \timeform{02h17m30.04s} & \timeform{-5D24'31.6''} & 1.429 & $8.66_{-0.02}^{+0.02}$ & $1.1\times10^{11}$ 
& 233 & 1006 \\
\hline

\multicolumn{1}{@{}l@{}}{\hbox to 0pt{\parbox{160mm}
{\footnotesize
\footnotemark[$\ast$] We adopted Salpeter IMF. \\
\footnotemark[$\dagger$] SFRs are derived from extinction corrected UV luminosity densities. \\
\footnotemark[$\ddagger$] SFRs are calculated by summing the SFRs derived from 24 $\mu$m flux densities and derived from UV luminosity densities. \\
\footnotemark[$\S$] Galaxy sample in \citet{Seko14}. Since COSMOS\_9 especially showed a signal-like feature, 
we try to detect the CO emission again.
}
\hss}}
\end{longtable}

\section{Sample} \label{sec: sample}
The galaxy sample was taken from \authorcite{Yabe12}~(\yearcite{Yabe12}, \yearcite{Yabe14}) and \citet{Rose12}. 
They conducted near-infrared spectroscopic observations toward star-forming galaxies at $z\sim1.4$ 
in the Subaru XMM/Newton Deep Survey (SXDS) field 
and in the Cosmological Evolution Survey (COSMOS) field, respectively, 
using Fiber Multi Object Spectrograph (FMOS) on the Subaru telescope. 
The galaxy sample by \authorcite{Yabe12}~(\yearcite{Yabe12}, \yearcite{Yabe14}) 
consists of {\it K}-band selected galaxies that are mostly on the main sequence at $z_\mathrm{phot}\sim1.4$. 
On the other hand, the galaxy sample in \citet{Rose12} consists of MIPS and SPIRE selected sources. 
The spectroscopic redshift was derived using H$\alpha$ emission line. 
The gas phase metallicity was derived with N2 method by using the H$\alpha$ and 
[NII]$\lambda$~6584 emission lines \citep{PP04}. 
We selected the galaxies with the metallicity of $12+\log(\mathrm{O/H})>8.6$ 
to reduce the uncertainty of CO-to-H$_2$ conversion factors. 

In order to derive the far-infrared luminosity and the dust mass, we further required 
the galaxies are detected with MIPS in $24~\mu\mathrm{m}$, 
SPIRE in $250~\mu\mathrm{m}$ and $350~\mu\mathrm{m}$. 
We selected sources that appear isolated in the $24~\mu\mathrm{m}$ image, 
allowing us to obtain reliable flux density in mid/far-infrared. 
In fact, the contamination of target galaxies by adjacent sources in the images 
at $250~\mu\mathrm{m}$ and $350~\mu\mathrm{m}$ is not serious. 
For the galaxies in the SXDS field, MIPS data were taken from the DR2 version of 
{\it Spitzer} Public Legacy Survey of the UKIDSS (UKIRT Infrared Deep Sky Survey) 
Ultra Deep Survey (SpUDS: J. Dunlop et al. in preparation). 
SPIRE data were taken from the DR1 version of the Herschel Multi-tired Extragalactic Survey 
(HerMES: \cite{Oliv12}). 
Object detection and photometry were made with SExtractor \citep{Bert96}. 
For the galaxies in the COSMOS field, the photometric data of MIPS and SPIRE were taken from \citet{Rose12}. 
We made SED from MIR to FIR, and derived $L_\mathrm{FIR}(8-1000~\mu\mathrm{m})$ 
by fitting model SEDs of star-forming galaxies \citep{CE01}. 

\begin{figure*}[!H]
\begin{center}
\includegraphics[width=17cm]{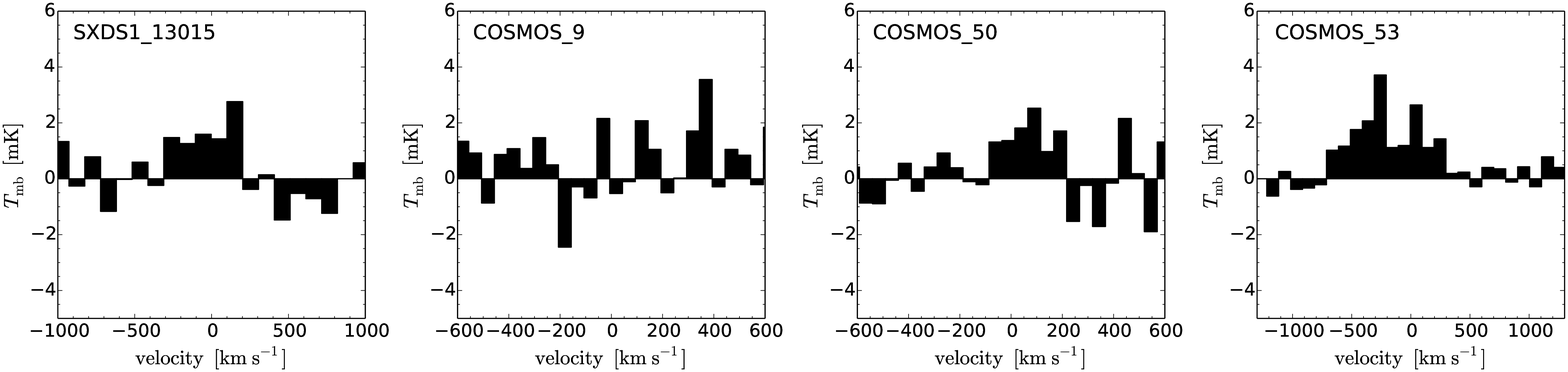} 
\end{center}
\caption{CO($2-1$) spectra of SXDS1\_13015 (left), COSMOS\_9 (center left), COSMOS\_50 (center right), 
	and COSMOS\_53 (right) taken with the Nobeyama 45~m telescope. 
	Spectra are binned with a $50~\mathrm{km~s^{-1}}$ velocity width for COSMOS\_9 and COSMOS\_50, 
	with a $100~\mathrm{km~s^{-1}}$ velocity width for COSMOS\_53 and SXDS1\_13015. 
	The zero velocity is derived by referring the spectroscopic redshift by H$\alpha$ observation. 
	} \label{fig: spectrum}
\end{figure*}

Among these selected galaxies, we chose galaxies that are located around the main-sequence 
of star-forming galaxies \citep{Dadd10}. 
For the galaxies in the SXDS field, the stellar mass and SFR were taken from 
\authorcite{Yabe12}~(\yearcite{Yabe12}, \yearcite{Yabe14}); 
the stellar masses are derived by fitting the SED from far-UV to mid-IR data 
and SFRs are derived from the rest-frame UV luminosity densities corrected 
for the dust extinction estimated from the rest-frame UV-slope. 
For the galaxies in the COSMOS field, using the photometry data in \citet{Muzz13}, 
we derived stellar mass and SFR with the same method as those in the SXDS field. 
We selected four galaxies and they are shown in Table~\ref{tab: sample}~\footnote{
COSMOS\_9 was already observed by \citet{Seko14}, but its spectrum only exhibited a 
tentative CO-detection. Thus, we decided to re-observe COSMOS\_9. 
In \citet{Seko14}, the stellar mass and SFR of COSMOS\_9 were taken from \citet{Whit11}. 
The stellar mass and SFR are 3.7 times larger and 2.3 times smaller 
than those in this paper, respectively.}. 
The properties in mid and far-infrared are also listed in Table~\ref{tab: sample dust}. 
In Table~\ref{tab: sample}, SFR derived from extinction corrected UV luminosity density 
and from the sum of the $24~\mu\mathrm{m}$ luminosity density and extinction uncorrected 
UV luminosity density, 
and they are shown in left panel and right panel of Fig.~\ref{fig: sample}, respectively (red circles). 
For the sample in this paper (except for COSMOS\_9), the SFRs from $24~\mu\mathrm{m}$ luminosity density 
and extinction uncorrected UV luminosity density are consistent with those 
from extinction corrected UV luminosity density. 
The sample by \citet{Seko14} is also shown in Table~\ref{tab: sample} and \ref{tab: sample dust}
and Fig.~\ref{fig: sample} (black circles) together with the sample in this paper. 

\begin{table}
\tbl{Mid and far-infrared properties of galaxy sample. }{%
\begin{tabular}{lcccc}
\hline
Source & $S_{24}$ & $S_{250}$ & $S_{350}$ & $L_\mathrm{FIR}$ \\
 & (mJy) & (mJy) & (mJy) & ($L_\odot$) \\
\hline
SXDS1\_13015 & $0.32\pm0.02$ & $43\pm8$ & $44\pm10$ & $3.4\times10^{12}$ \\
COSMOS\_9     & $0.44\pm0.01$ & $26\pm3$ & $25\pm3$ & $4.2\times10^{12}$ \\
COSMOS\_50   & $0.76\pm0.01$ & $47\pm3$ & $34\pm3$ & $2.9\times10^{12}$ \\
COSMOS\_53   & $0.36\pm0.01$ & $26\pm3$ & $22\pm3$ & $1.1\times10^{12}$ \\
\hline
SXDS1\_12778 & $0.46\pm0.02$ & $45\pm7$ & $59\pm8$ & $3.8\times10^{12}$ \\
SXDS3\_80799 & $0.59\pm0.02$ & $54\pm6$ & $46\pm5$ & $5.9\times10^{12}$ \\
\hline
\end{tabular}} \label{tab: sample dust}
\end{table}


\section{CO($J$=2-1) observations and data reduction} \label{sec: obs and reduc}
We made $^{12}$CO($J=2-1$) line observations toward the four star-forming galaxies at $z\sim1.4$ 
on 2014 March 22nd, 23rd, 25th (COSMOS\_9 and COSMOS\_50), 
2015 January 29th, 31st, and February 1st (COSMOS\_53 and SXDS1\_13015) 
using the Nobeyama 45~m telescope. 
The observing frequencies were 93.677 to 104.316~GHz 
calculated with spectroscopic redshifts derived from the near-infrared observations. 
We used the two-beam sideband-separating SIS receiver for $z$-machine with dual polarization (TZ; \cite{Naka13}). 
The half power beam width at these frequencies was $\sim17$~arcsec. 
We used the flexible FX-type spectrometer (spectral analysis machine for the 45~m telescope (SAM45); \cite{Iono12}). 
We can use up to 16 spectral windows (SPWs) 
and choose a band width of each SPW from several modes between 16~MHz and 2~GHz. 
To cover the wider range of velocity, we selected the 2~GHz mode.
The frequency width of one channel is 488.28~kHz, because each array has 4096 channels. 
The image rejection ratios in the observing frequency ranges exceeded 10~dB.
The system noise temperature ($T_\mathrm{sys}$) was typically 130-170~K.
The accuracy of telescope pointing was checked every 50~min with the observations of SiO maser sources 
($o$~Cet and R~Leo) near the galaxy sample.
During the observations, the accuracy was within $3$~arcsec.

We used the NEWSTAR software for the data reduction. 
The data taken under the condition of a wind speed of less than $5~\mathrm{m~s^{-1}}$ were used. 
In addition, we flagged data with poor baseline by visual inspection.
Three persons independently set three criteria for the flagging to check the robustness of the results. 
All analyses showed similar results. 
After flagging, the effective integration time per galaxy was 4.2-8.8~hours. 
All data were converted from the antenna temperature ($T_\mathrm{A}$) to main beam temperature ($T_\mathrm{mb}$). 
The main beam efficiencies were 0.38 for COSMOS\_9 and COSMOS\_50 
and of 0.42 for COSMOS\_53 and SXDS1\_13015~\footnote{
If we adopted an aperture efficiency of 0.26 for COSMOS\_9/COSMOS\_50 and 0.27 for COSMOS\_53/SXDS1\_13015, 
their molecular gas masses and gas-to-dust ratios are 1.5 and 1.6 larger than those estimates with the main beam efficiency, 
respectively.}. 
The root mean square noise temperature in $T_\mathrm{mb}$ scale was 0.7-1.5~mK. 

\begin{longtable}{lccccc}
\caption{Molecular gas and dust properties of galaxy sample.} \label{tab: gas and dust}
\hline              
Source & $L_\mathrm{CO(1-0)}^{'}$ & $M_\mathrm{mol}$\footnotemark[$\ast$] & $f_\mathrm{mol}$ & $M_\mathrm{d}$ & Gas-to-dust ratio \\ 
 & ($\mathrm{K\ km\ s^{-1}\ pc^2}$) & ($M_\odot$) &  & ($M_\odot$) & \\
\endhead
  \hline
SXDS1\_13015 & $(5.4\pm1.3)\times10^{10}$ & $(2.4\pm0.6)\times10^{11}$ & $0.54\pm0.07$ & $5.4\times10^{8}$ & $440\pm110$ \\
COSMOS\_9     & $<2.2\times10^{10}$ & $<9.7\times10^{10}$ & $<0.59$ & $3.4\times10^{8}$ & $<290$ \\
COSMOS\_50   & $(2.2\pm0.5)\times10^{10}$ & $(9.6\pm2.1)\times10^{10}$ & $0.58\pm0.07$ & $4.5\times10^{8}$ & $220\pm50$ \\
COSMOS\_53   & $(8.1\pm0.8)\times10^{10}$ & $(3.5\pm0.3)\times10^{11}$ & $0.78\pm0.02$ & $2.4\times10^{8}$ & $1450\pm140$  \\
			 &  & $(6.5\pm0.6)\times10^{10}$\footnotemark[$\dagger$] & $0.40\pm0.01$\footnotemark[$\dagger$] &  & $265\pm25$\footnotemark[$\dagger$] \\
\hline
SXDS1\_12778 & $<2.9\times10^{10}$ & $<1.3\times10^{11}$ & $<0.19$ & $5.4\times10^{8}$ & $<240$ \\
SXDS3\_80799 & $<2.2\times10^{10}$ & $<9.7\times10^{10}$ & $<0.47$ & $6.7\times10^{8}$ & $<150$ \\
\hline
\multicolumn{1}{@{}l@{}}{\hbox to 0pt{\parbox{160mm}
{\footnotesize
\footnotemark[$\ast$] We adopted $\alpha_\mathrm{CO} = 4.36~M_\odot~\mathrm{(K~km~s^{-1}~pc^2)^{-1}}$. \\
\footnotemark[$\dagger$] We adopted $\alpha_\mathrm{CO} = 0.8~M_\odot~\mathrm{(K~km~s^{-1}~pc^2)^{-1}}$.
}
\hss}}
\end{longtable}


\begin{figure*}
\begin{center}
\includegraphics[width=17cm]{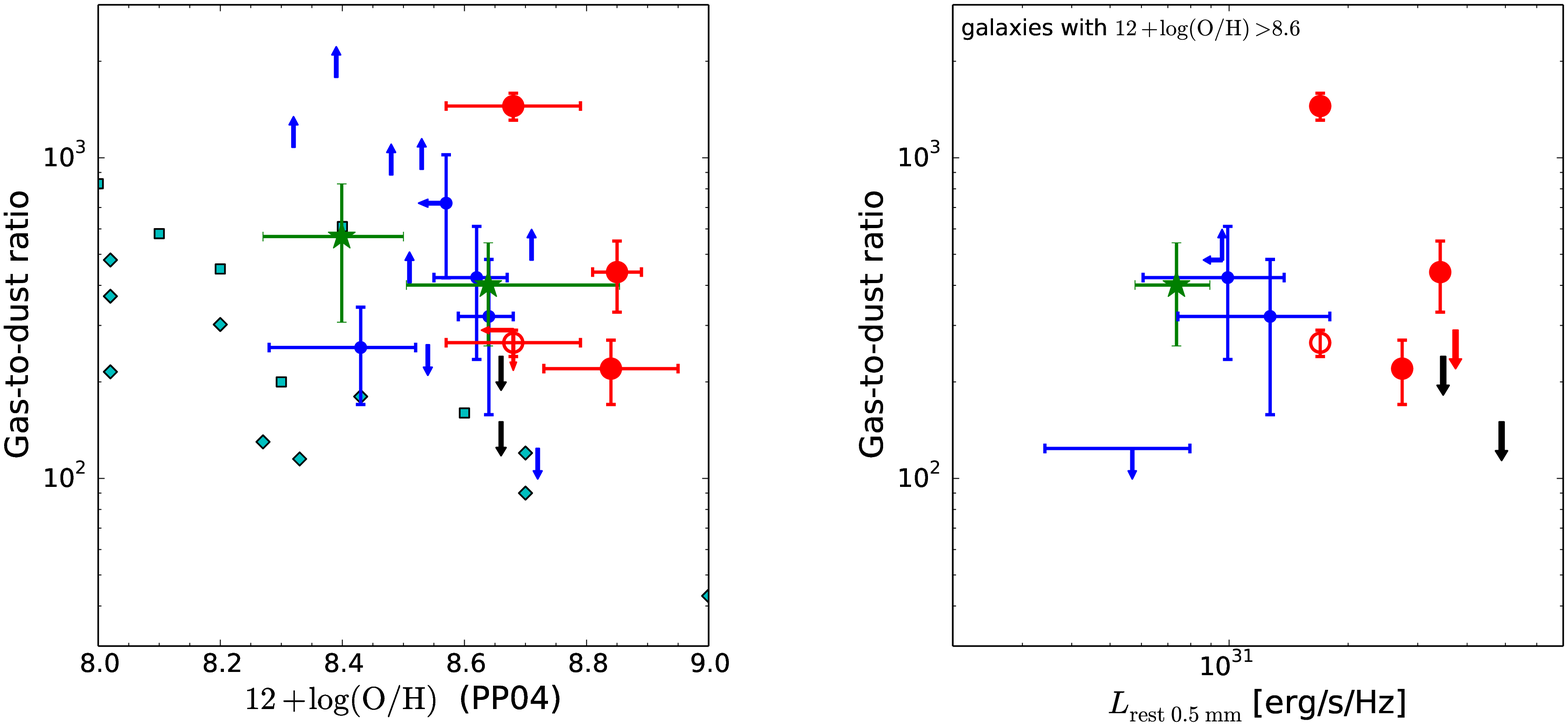} 
\end{center}
\caption{Gas-to-dust ratio against metallicity (left), and luminosity density at rest-wavelength of 0.5~mm 
	for galaxies with $12+\log(\mathrm{O/H}) > 8.6$ (right). 
	Red and black symbols refer to the galaxies in this paper and in \citet{Seko14}, respectively. 
	Open circle represents COSMOS\_53 with the ULIRG-like $\alpha_\mathrm{CO}$. 
	Blue filled circles refer to the individual galaxies and green stars refer 
	to the results of metallicity-based stacking analysis by \citet{Seko16}. 
	Cyan diamonds represent local galaxies by \citet{Lero11} and 
	cyan squares represent the average values in local galaxies shown by \citet{Remy14}. 
	(Metallicities are calibrated using \cite{PP04}.) 
} \label{fig: GDR}
\end{figure*}

\section{Results} \label{sec: result}
\subsection{CO(2-1) spectra}
The spectra obtained are shown in Fig.~\ref{fig: spectrum}. 
The CO($2-1$) emission line was detected toward SXDS1\_13015, COSMOS\_50, and COSMOS\_53 
but was not detected toward COSMOS\_9. 
The spectrum of SXDS1\_13015 shows velocity width of $500~\mathrm{km~s^{-1}}$, 
and the noise level at a velocity resolution of $500~\mathrm{km~s^{-1}}$ ($\sigma_{500}$) is 0.41~mK. 
The signal-to-noise ratio (S/N) of SXDS1\_13015 at the $500~\mathrm{km~s^{-1}}$ resolution is 4.1. 
The spectrum of COSMOS\_50 shows a velocity width of 300~$\mathrm{km~s^{-1}}$, 
and the noise level at the velocity resolution ($\sigma_{300}$) is 0.35~mK. 
The S/N of COSMOS\_50 at the $300~\mathrm{km~s^{-1}}$ resolution is 4.6. 
The spectrum of COSMOS\_53 shows a velocity width of $1000~\mathrm{km~s^{-1}}$, 
and the noise level at the velocity resolution ($\sigma_{1000}$) is 0.16~mK. 
The S/N of COSMOS\_53 at the $1000~\mathrm{km~s^{-1}}$ resolution is 10.5. 
The velocity width is larger than a typical value ($\sim200~\mathrm{km~s^{-1}}$) 
in main sequence galaxies at a similar redshift (e.g., \cite{Tacc13, Seko16}). 
We checked an $I_{F814W}$-band image taken with the Advanced Camera for Surveys (ACS) 
on the {\it Hubble Space Telescope} \citep{Koek07}. 
The image of COSMOS\_53 shows that the target may be a merging system, 
thus the CO emission line may include both merging galaxies.

\subsection{Molecular gas mass}
We calculated the CO($1-0$) luminosity ($L_\mathrm{CO(1-0)}^{'}$) 
from the obtained spectrum with the following equation:
\begin{equation}\label{Lco}
L_\mathrm{CO(1-0)}^{'} = 3.25 \times 10^7\ S_\mathrm{CO(2-1)} \Delta v R_{21}^{-1} \nu_\mathrm{rest(1-0)}^{-2} D_L^2 (1 + z)^{-1},
\end{equation}
where $L_\mathrm{CO(1-0)}^{'}$ is measured in $\mathrm{K~km~s^{-1}~pc^2}$, 
$S_\mathrm{CO(2-1)}$ is the observed CO$(2-1)$ flux density in Jy, 
$\Delta v$ is the velocity width in $\mathrm{km~s^{-1}}$, 
$R_{21}$ is the flux ratio of CO$(2-1)$ to CO$(1-0)$, 
$\nu_\mathrm{rest(1-0)}$ is the rest frequency of the CO(1-0) emission line in GHz, 
and $D_L$ is the luminosity distance in Mpc. 
The values of $\Delta v$ for SXDS1\_13015, COSMOS\_50, and COSMOS\_53 
are $500~\mathrm{km~s^{-1}}$, $300~\mathrm{km~s^{-1}}$, and $1000~\mathrm{km~s^{-1}}$, respectively.
The value of $R_{21}$ is assumed to be 3, which is a typical value 
for color selected star-forming galaxies at $z=1-3$ (\cite{Cari13, Dadd15}). 
For COSMOS\_9, we derived the $2\sigma_{250}$ upper limit of the CO($1-0$) luminosity 
($\sigma_{250}$ is a noise level at a 250~$\mathrm{km~s^{-1}}$ velocity resolution) 
assuming a velocity width of $250~\mathrm{km~s^{-1}}$, as in \citet{Seko14}. 
The derived CO($1-0$) luminosities of the galaxies are shown in Table~\ref{tab: gas and dust}. 

Then, we derived the molecular gas mass by using the CO-to-H$_2$ conversion factor ($\alpha_\mathrm{CO}$). 
The $\alpha_\mathrm{CO}$ correlates with metallicity in local star-forming galaxies; 
the value of $\alpha_\mathrm{CO}$ is smaller in galaxies with higher metallicity (e.g., \cite{Arim96,Lero11}). 
According to recent studies, a similar relation is seen in star-forming galaxies at $z=1-2$ \citep{Genz12}, 
and $\alpha_\mathrm{CO}$ is close to the Galactic value for the main sequence star-forming galaxies 
with solar metallicity (e.g., \cite{Dadd10, Genz12}). 
The uncertainty in $\alpha_\mathrm{CO}$ is smaller than that at lower metallicity. 
Since the galaxies selected have metallicity close to the solar metallicity, 
we adopt the Galactic $\alpha_\mathrm{CO}$ value of 
$4.36~M_\odot~(\mathrm{K~km~s^{-1}~pc^2})^{-1}$ (including helium mass). 
For COSMOS\_53, since it is possible that this galaxy is undergoing a merging, 
we also derived the molecular gas mass adopting the ULIRG-like conversion factor 
(0.8~$M_\odot~(\mathrm{K~km~s^{-1}~pc^2})^{-1}$; \cite{Down98}). 
The resulting molecular gas masses except for COSMOS\_9 are $(1-3.5)\times10^{11}~M_\odot$ as given in Table~\ref{tab: gas and dust}. 
The uncertainty given to the molecular gas mass is based on SN of the spectrum 
and does not include the uncertainty of the luminosity ratio and $\alpha_\mathrm{CO}$. 
The fractions of molecular gas against stellar mass are 0.5-0.8 
(for COSMOS\_53, $f_\mathrm{mol}=0.4$ with the ULIRG-like conversion factor).
These values are in agreement with those in the previous studies (e.g., \cite{Dadd10, Tacc10, Tacc13, Seko16}).

\subsection{Dust mass}
The dust mass is derived from 
\begin{equation}
\label{eq: dust mass}
M_\mathrm{d} = \frac{S_\mathrm{obs} D_L^2}{(1 + z) \kappa_\mathrm{d} (\nu_\mathrm{rest}) B (\nu_\mathrm{rest}, T_\mathrm{d})},
\end{equation}
where $S_\mathrm{obs}$ is the observed flux density, 
$\kappa_\mathrm{d} (\nu_\mathrm{rest})$ is the dust mass absorption coefficient in the rest-frame frequency, 
$T_\mathrm{d}$ is the dust temperature, and $B (\nu_\mathrm{rest}, T_\mathrm{d})$ is the Planck function. 
$\kappa_\mathrm{d}$ varies with frequency as $\kappa_\mathrm{d} \propto \nu^{\beta}$, 
where $\beta$ is the dust emissivity index. 
As in \citet{Seko14}, 
we used a $250~\mu\mathrm{m}$ flux density for $S_\mathrm{obs}$, 
$\kappa_\mathrm{d} (125\ \mu \mathrm{m}) = 1.875\ \mathrm{m^2\ kg^{-1}}$ \citep{Hild83}, 
$\beta=1.5$, and $T_\mathrm{d}=35~\mathrm{K}$. 
Since our galaxy sample tends to be located in the upper part of the main sequence, 
we adopted a typical dust temperature for galaxies with higher specific SFRs among main sequence galaxies \citep{Magn14}. 
The dust mass can change by a factor of $\sim$1.8 when we adopt dust temperatures of 30 or 40~K. 
The uncertainty arisen from $\beta$ is 10~\% in adopting $\beta=$1.0 or 2.0, 
since we use the flux density at the rest-wavelength of 105~$\mu\mathrm{m}$ 
and we use $\kappa_\mathrm{d}$ at 125~$\mu\mathrm{m}$. 
The resulting dust masses are $(2.4-5.4)\times10^{8}~M_\odot$ as given in Table~\ref{tab: gas and dust}. 

\subsection{Gas-to-dust ratio}
We calculated gas-to-dust ratios ($M_\mathrm{mol}/M_\mathrm{dust}$) of the galaxy sample. 
The resulting gas-to-dust ratios are $220-1450$ as shown in Table~\ref{tab: gas and dust} 
and plotted against the gas metallicity with red circles and the arrow (upper limit) in the left panel of Fig.~\ref{fig: GDR}. 
For COSMOS\_53, the gas-to-dust ratio is 265 (open red circle), 
when we use the ULIRG-like CO-to-H$_2$ conversion factor. 
Note that the gas-to-dust ratio does not include the mass of the atomic hydrogen gas (HI). 
According to a semi-empirical model by \citet{Popp15}, HI mass in galaxies at $z\sim1.4$ 
with halo mass of $10^{12-14}~M_\odot$ is half or comparable to the H$_2$ mass. 
Therefore, the actual gas-to-dust ratios may be larger by 0.2-0.3 dex. 

\section{Discussion} \label{sec: dis}
In the left panel of Fig.~\ref{fig: GDR}, we also plot the upper limits of the gas-to-dust ratio at $z\sim1.4$ 
by \citet{Seko14} (SXDS1\_12778 and SXDS3\_80799) with black arrows. 
The star-forming galaxies with solar metallicity detected with mid/far-infrared (MIPS and SPIRE) 
show various gas-to-dust ratios. 
The ratios in local galaxies (including HI mass) are also shown with cyan diamonds (\cite{Lero11}) 
and squares (\cite{Remy14}). 
The ratios in SXDS1\_13015, COSMOS\_50, and COSMOS\_53 (even if we use the ULIRG-like CO-to-H$_2$
conversion factor for COSMOS\_53) are several times larger than those in local galaxies with solar metallicity. 
While dust masses are derived from the modified blackbody model in this study, 
if we adopt the dust model by \citet{DL07} (hereafter DL07 model), the dust mass is systematically 2-3 times larger than 
that derived from the modified blackbody (e.g., \cite{Magd12b, Magn12}). 
If we used the DL07 model for deriving dust mass, our results can shift to near the local values. 
As mentioned above, if the HI mass is included, the gas-to-dust ratio 
at $z$$\sim$1.4 is still larger than that in the local universe.

\citet{Seko16} conducted observations of $^{12}$CO($J=5-4$) emission line and dust continuum emission 
toward twenty star-forming galaxies at $z\sim1.4$ with wider range of stellar mass and metallicity by using ALMA.  
Almost all of their galaxies were not detected with MIPS nor SPIRE. 
They found that the gas-to-dust ratio in main sequence galaxies at $z\sim1.4$ is 
3-4 times larger than that in local galaxies at each fixed metallicity. 
In the left panel of Fig.~\ref{fig: GDR}, the result of stacking analysis by \citet{Seko16} is plotted with green stars. 
They adopted the metallicity-dependent CO-to-H$_2$ conversion factor by \citet{Genz12}. 
In addition, they assumed $T_\mathrm{d}=30~\mathrm{K}$, because their galaxies are located around
the center of main sequence by \citet{Dadd10}. 
If $T_\mathrm{d}=35~\mathrm{K}$ is assumed, the gas-to-dust ratios in \citet{Seko16} is larger 
than those with $T_\mathrm{d}=30~\mathrm{K}$ by $\sim0.1$~dex. 
Results in this study are roughly consistent with those by \citet{Seko16}. 
One galaxy, SXDS1\_13015, was also observed by \citet{Seko16}. 
By adopting the same conversion factor and dust temperature in this paper, 
the gas-to-dust ratio derived by \citet{Seko16} ($\sim$470; $M_\mathrm{mol}=1.8\times10^{11}~M_\odot$ 
and $M_\mathrm{dust}=3.8\times10^{8}~M_\odot$)
is in agreement with that obtained in this paper.

To examine the dependence of gas-to-dust ratio on far-infrared luminosity density, 
we plot the gas-to-dust ratios in the star-forming galaxies with solar metallicity ($12+\log(\mathrm{O/H})>8.6$)
in this study and those by the previous studies (\cite{Seko14}, \yearcite{Seko16}) 
against the rest-frame luminosity density at 0.5~mm ($L_\mathrm{rest~0.5~mm}$; right panel of Fig.~\ref{fig: GDR}). 
Although it may be better to see the dependence on the total infrared luminosity rather than the luminosity density, 
since we only have the flux densities in the band-6 for almost all of the ALMA sample 
and the total luminosity depends on the SED, we use the $L_\mathrm{rest~0.5~mm}$ to avoid the uncertainty. 
The luminosity densities of the galaxies in this study and \citet{Seko14} are estimated 
from the best fitted model SED by \citet{CE01}. 
The gas-to-dust ratio does not seem to depend on the luminosity density at $0.5~\mathrm{mm}$. 
Although most of the ratios are ~200-500 (if we use the ULIRG-like conversion factor for COSMOS\_53), 
the lower and upper limits indicate that star-forming galaxies with solar metallicity at $z\sim1.4$
probe a wide range of gas-to-dust ratios.

\section{Summary}\label{sec: sum}
We carried out $^{12}\mathrm{CO}(J=2-1)$ observations 
toward four massive star-forming galaxies at $z\sim1.4$ around the main sequence
with the Nobeyama 45~m radio telescope. 
The galaxies were detected with {\it Spitzer}/MIPS in 24~$\mu\mathrm{m}$ 
and {\it Herschel}/SPIRE in 250~$\mu\mathrm{m}$, and 350~$\mu\mathrm{m}$.  
The metallicities of these galaxies which were derived with N2 method were near the solar value. 
CO emission lines were detected toward three of the galaxies. 
The masses of molecular gas are $(9.6-35)\times10^{10}~M_\odot$ 
by adopting the Galactic CO-to-H$_2$ conversion factor and the CO(2-1)/CO(1-0) flux ratio of 3. 
The masses of dust are  $(2.4-5.4)\times10^{8}~M_\odot$ 
with the modified blackbody model by assuming 
$T_\mathrm{dust}=35~\mathrm{K}$ and $\beta=1.5$. 
The resulting gas-to-dust ratios (not accounting for HI) are $220-1450$. 
Most of them are several times larger than those in local galaxies with solar metallicity. 
The dependence of the gas-to-dust ratio on the far-infrared luminosity density is not clearly seen, 
even if we include the galaxy sample in \citet{Seko16} whose luminosity density is smaller than 
those of the galaxy sample in this paper.

\begin{ack}
We would like to thank the referee for useful comments and suggestions.
We acknowledge the members of Nobeyama Radio Observatory for their help during the observations. 
A.S. is supported by Research Fellowship for Young Scientists 
from the Japan Society of the Promotion of Science (JSPS). 
K.O. was supported by Grant-in-Aid for Scientific Research (C) (24540230) from JSPS. 
Kavli IPMU is supported by World Premier International Research Center Initiative (WPI), MEXT, Japan.
\end{ack}


\end{document}